\documentclass[amsmath,showpacs,pre]{revtex4}
\usepackage{graphicx}
\usepackage{amsmath}
\usepackage{amssymb}
\usepackage{natbib}
\usepackage{subfigure}
\usepackage{color} 

\begin{document}

\title{Vibrations of Jammed Disk Packings with Hertzian 
Interactions}

\author{Carl F. Schreck$^{1,2}$}
\author{Corey S. O'Hern$^{1,2,3}$}
\author{Mark D. Shattuck$^{4,1}$} 
 
\affiliation{$^{1}$Department of Mechanical Engineering \& Materials Science, Yale University, New Haven, Connecticut 06520-8260, USA}

\affiliation{$^{2}$Department of Physics, Yale University, New Haven,
Connecticut 06520-8120, USA}

\affiliation{$^{3}$Department of Applied Physics, Yale University, New Haven, Connecticut 06520-8120, USA}

\affiliation{$^{4}$Benjamin Levich Institute and Physics Department, The City College of the City University of New York, New York, New York 10031, USA}

\begin{abstract}
Contact breaking and Hertzian interactions between grains can both give
rise to nonlinear vibrational response of static granular
packings. We perform molecular dynamics simulations at constant energy
in 2D of frictionless bidisperse disks that interact via Hertzian
spring potentials as a function of energy and measure {\it directly}
the vibrational response from the Fourier transform of the velocity
autocorrelation function.  We compare the measured vibrational
response of static packings near jamming onset to that obtained from
the eigenvalues of the dynamical matrix to determine the temperature
above which the linear response breaks down.  We compare packings that
interact via single-sided (purely repulsive) and double-sided Hertzian
spring interactions to disentangle the effects of the shape of the
potential from contact breaking. Our studies show that while Hertzian
interactions lead to weak nonlinearities in the vibrational behavior
(e.g. the generation of harmonics of the eigenfrequencies of the
dynamical matrix), the vibrational response of static packings with
Hertzian contact interactions is dominated by contact breaking as found for
systems with repulsive linear spring interactions.
\end{abstract}

\pacs{
63.50.Lm, 
63.50.-x, 
64.70.pv, 
83.80.Fg
}

\maketitle

\section{Introduction}
\label{intro}

Dry granular media are composed of discrete grains that interact via
purely repulsive, frictional contact interactions.  Without external
driving, granular materials form static packings that possess
nonlinear response to external perturbations~\cite{makse,wildenberg}.
For example, the acoustic response of granular packings includes
harmonic mode generation, mixing, attenuation, and dispersion, which
have been exploited to engineer novel phononic metamaterials that can
act as rectifiers and filters~\cite{boechler}.  There are many sources
of nonlinearity in granular media, which include 1) the nonlinear form
of Hertzian interactions between grains~\cite{johnson}, 2) contact
breaking and formation ({\it i.e.}  contact clapping~\cite{tournat})
that occurs frequently in systems with purely repulsive contact
potentials, and 3) rolling and sliding frictional contacts. In prior
studies, we isolated the nonlinearities that arise from contact
breaking by measuring the vibrational response of mechanically stable
(MS) packings of frictionless disks that interact via purely repulsive
linear spring potentials~\cite{tibo1,tibo2}. In this manuscript, we
perform computational studies to determine the relative contributions
of the nonlinearities that arise from the shape of the Hertzian
potential and contact breaking by measuring the vibrational response
of MS packings that interact via single- (repulsive only) and
double-sided (repulsive and attractive) Hertzian springs.

We find two overarching results: 1) The shape of the Hertzian
interaction potential gives rise to nonlinearities in the vibrational
response of jammed disk packings, such as the generation of harmonics
of the driving frequency and beats among these and normal mode
frequencies from the dynamical matrix. 2) However, these
nonlinearities are weak compared to those generated by contact
breaking in systems with purely repulsive Hertzian spring
interactions.  In particular, prior to contact breaking (over the
timescales we considered), the measured density of vibrational modes
for jammed packings with Hertzian interactions is similar to that
inferred from linear response.  These results emphasize the importance
of contact breaking in determining the vibrational response in jammed
packings that interact via purely repulsive linear spring as well as
Hertzian potentials.

\section{Methods}
\label{methods}

We first prepared MS packings of frictionless disks using the
successive-compression-and-decompression algorithm~\cite{gao} at a given
deviation in packing fraction above jamming onset $\Delta \phi = \phi
- \phi_J$ for systems composed of $N=32$ to $128$ disks.  We then 
performed molecular dynamics simulations at constant total energy $E$
in a periodic square cell of $N$ frictionless
disks that interact via purely repulsive pair potentials:
\begin{eqnarray}
\label{spring}
{\vec F}_{ij}&=&
\frac{\epsilon}{\sigma_{ij}}\bigg(1-\frac{r_{ij}}{\sigma_{ij}}\bigg)^{\alpha}\Theta\bigg(1-\frac{r_{ij}}{\sigma_{ij}}\bigg)\hat{r}_{ij}
\end{eqnarray}
where $r_{ij}$ is the separation between particles $i$ and $j$, ${\hat
r}_{ij}=(x_{ij}{\hat x} + y_{ij} {\hat
y})/\sqrt{x^2_{ij}+y^2_{ij}}$ is a unit vector that points from
particle $j$ to $i$, $\Theta(1-r_{ij}/\sigma_{ij})$ is the Heaviside
step function that ensures that particles do not interact when they do
not overlap, $\sigma_{ij}=(\sigma_i+\sigma_j)/2$, $\sigma_i$ is the
diameter of disk $i$, and $\epsilon=1$ is the characteristic energy
scale of the repulsive interaction.  The power-law exponent $\alpha$
determines the form of the repulsive interactions, where $\alpha = 1$
($3/2$) denotes the linear (Hertzian) spring interaction.  We consider
bidisperse mixtures with half large and half small particles and size
ratio $d=1.4$ to inhibit crystallization~\cite{leo}.

The linear vibrational response for MS packings (that interact via
the pairwise forces in Eq.~\ref{spring}) can be obtained from the dynamical
matrix~\cite{barrat}
\begin{equation}
\label{m}
K_{\alpha \beta} = \left. \frac{d^2 V}{d\xi_{\alpha} d\xi_{\beta}}\right|_{{\vec \xi}={\vec \xi}^0},
\end{equation}
where $V=\sum_{i,j=1}^N V(r_{ij})$ is
the total potential energy, ${\vec F}_{ij} = -dV(r_{ij})/dr_{ij} {\hat
r}_{ij}$, ${\vec \xi}=\{x_1,y_1,x_2,y_2,\ldots,x_N,y_N\}$ give the
positions of the disk centers, and $K$ is evaluated at the MS packing,
${\vec \xi}^0$, which is a local minimum of $V$. $K$ is a $2N\times 2N$ matrix that 
can be written in terms of the $2\times 2$ block matrices
\begin{equation}
K'_{lm} =
\bigg(\begin{array}{cc}
K_{x_lx_m} & K_{x_ly_m} \\
K_{y_lx_m} & K_{y_ly_m} \\
\end{array}\bigg),
\label{k}
\end{equation}
which in the $\Delta \phi\rightarrow 0$ limit reduces to
\begin{equation}
K'_{lm} \approx \alpha
\bigg(\frac{\epsilon}{\sigma_{lm}^4}\bigg)
\bigg(\frac{\sigma_{lm}F_{lm}}{\epsilon}\bigg)^{(\alpha-1)/\alpha}
\bigg(\begin{array}{cc}
x_{lm}^2 & x_{lm}y_{lm} \\
x_{lm}y_{lm} & y_{lm}^2 \\
\end{array}\bigg),
\label{kprime}
\end{equation}
where $l\ne m$, $l$,$m=1,\ldots,N$ are the disk labels, $x_{lm}=x_l-x_m$, 
$y_{lm}=y_l-y_m$, and $F_{lm}$ 
is the magnitude of the force ${\vec F}_{lm}$.  The $l$th
eigenvector of $K$, associated with eigenvalue $k_l$, is normalized so
that $\sum_{i=1}^{2N} e_{li}^2 = 1$, where ${\hat e_l} =
\{e_{lx_1},e_{ly_1},\ldots,e_{lx_N},e_{ly_N}\}$.  The vibrational
frequencies in the harmonic approximation are $\omega^d_l =
\sqrt{k_l/m}$, where $m$ is the mass of each disk.

We will compare the distribution of vibrational frequencies predicted
from linear response to the vibrational response measured in molecular
dynamics simulations. We vibrated each MS packing using two methods:
1) Perturb the static packing along one eigenmode $l$ of the dynamical
matrix by assigning the velocities $\vec{v}_i= \delta(E)\vec{e}_{li}$
to particle $i$, where ${\vec e}_{li}$ gives the components of the
$l$th eigenvector that correspond to particle $i$,
$\delta(E)=\sqrt{2E/m}$, and $E$ is the kinetic energy of the
perturbation; and 2) Perturb the system along a superposition of the
$2N-2$ eigenmodes~\cite{foot} by assigning ${\vec
v}_i=\delta(E)\vec{e}_{\rm all}$ where $\vec{e}_{\rm
all}=\frac{1}{\sqrt{2N-2}}\sum_{k=1}^{2N-2}\vec{e}_{ki}$.

For times $t>t_1=2\pi/\omega^d_1$ after the perturbation, where
$\omega^d_1$ is the smallest dynamical matrix frequency, we measure
the Fourier transform of the velocity autocorrelation function, which
yields the distribution of vibrational frequencies
\begin{equation}
\label{velocity}
D(\omega^v) = \int^{\infty}_0 dt \frac{\langle {\vec v}(t_0+t) \cdot {\vec v}(t_0)\rangle}{\langle {\vec v}(t_0)\cdot {\vec v}(t_0)\rangle} e^{i\omega^v t}, 
\end{equation}
where $\langle .\rangle$ indicates averages over all particles and
time origins $t_0$. We also measured the eigenvalue spectrum of the
displacement correlation matrix~\cite{dauchot}
\begin{equation}
\label{displacement}
C_{ij} = \langle (\xi_i -\xi_i^0) (\xi_j - \xi_j^0) \rangle,
\end{equation}
where $i,j=1\ldots 2N$ and the angle brackets indicate an average over
time.  The vibrational frequencies are obtained from the displacement
correlation matrix eigenvalues, $\omega_k^c = \sqrt{T/c_k}$. The
binned versions of the density of vibrational frequencies are given by
$D(\omega^{c,d})=[{\cal N}(\omega^{c,d}+\Delta \omega^{c,d})-{\cal
N}(\omega^{c,d})]/({\cal N}(\infty)\Delta \omega^{c,d})$, where ${\cal
N}(\omega)$ is the number of frequencies less than $\omega$.
$D(\omega^d)$, $D(\omega^v)$, and $D(\omega^c)$ are normalized so that
$\int_0^{\infty} d\omega D(\omega) = 1$.

\section{Results}
\label{results}

\begin{figure}[h!]
\begin{center}
\includegraphics[width=0.9\textwidth]{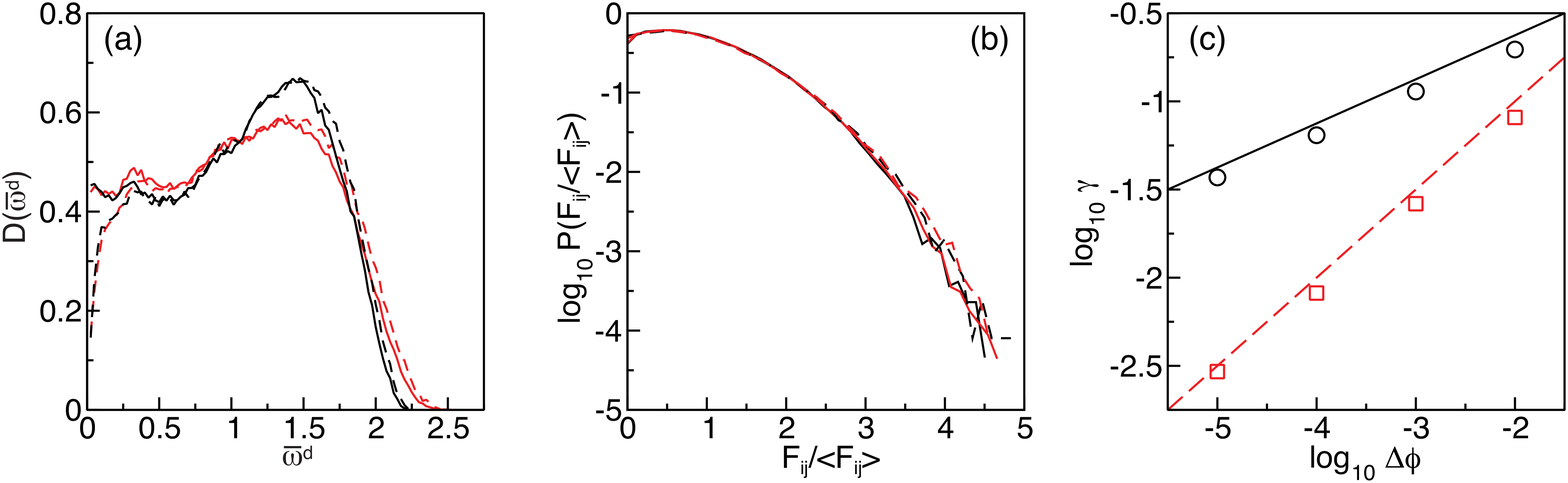}
\caption{(a) Density of vibrational modes $D({\overline \omega^d})$
from the dynamical matrix averaged over $800$ static packings with
$N=128$ disks that interact via purely repulsive linear (black lines)
and Hertzian (red lines) spring forces at $\Delta \phi=10^{-5}$ (solid
lines) and $\Delta \phi=10^{-2}$ (dashed lines). The contact networks
for the packings that interact via linear and Hertzian spring
interactions are identical. We define ${\overline \omega^d} =
\omega^d/(\langle \omega^d \rangle_{\alpha=1.5}/\langle \omega^d
\rangle_{\alpha=1})$, where $\langle \omega^d \rangle_{\alpha}$ is the
average vibration frequency for the force law in Eq.~\ref{spring} with
power-law exponent $\alpha$. (b) Distribution of normalized pair force
magnitudes $P(F_{ij}/\langle F_{ij} \rangle)$ for systems with purely
repulsive linear and Hertzian spring interactions at $\Delta \phi =
10^{-5}$ and $10^{-2}$ using the same line types in (a).  (c) Ratio of
the average vibrational frequencies for purely repulsive Hertzian and
linear spring interactions $\gamma_\omega=
\langle\omega^d\rangle_{\alpha=1.5}/\langle \omega^d
\rangle_{\alpha=1}$ (circles) and pair force magnitudes $\gamma_F=
\langle F_{ij}\rangle_{\alpha=1.5}/\langle F_{ij}\rangle_{\alpha=1}$
(squares) versus the deviation in packing fraction from jamming onset
$\Delta \phi$. The solid and dotted lines have slope $0.25$ and $0.5$,
respectively. }
\label{DOS_freqs_LS_Hertz_binned}
\end{center}
\end{figure}

In linear response, the density of vibrational modes is similar for
static packings that interact via purely repulsive linear and Hertzian
spring interactions.  In Fig.~\ref{DOS_freqs_LS_Hertz_binned} (a), we show
the density of vibrational modes $D(\overline \omega^d)$ obtained from
the dynamical matrix for linear and Hertzian springs.
Note that for Hertzian interactions, we considered scaled frequencies
${\overline \omega^d} = \omega^d/(\langle \omega^d
\rangle_{\alpha=1.5}/\langle \omega^d \rangle_{\alpha=1})$ since the
average frequency $\langle \omega^d \rangle_{\alpha=1.5} \sim (\Delta
\phi)^{0.25}$ for Hertzian springs as shown in
Fig.~\ref{DOS_freqs_LS_Hertz_binned} (c), whereas the average
vibrational frequency for linear spring potentials
$\langle\omega^d\rangle_{\alpha=1}$ is independent of $\Delta \phi$.

For packings near jamming onset that are
isostatic, the pair force magnitudes for systems that interact via
purely repulsive linear springs are proportional to those for Hertzian
springs as shown in Fig.~\ref{DOS_freqs_LS_Hertz_binned} (c).  From
Eqs.~\ref{k}, this implies that the dynamical matrix elements depend on
the force law, and thus the distribution of dynamical matrix
frequencies varies with the force law for isostatic packings near
jamming onset as emphasized in Fig.~\ref{DOS_freqs_LS_Hertz_binned} (a).

In Fig.~\ref{DOS_freqs_LS_Hertz_binned} (a), we show that the
distribution of vibrational frequencies from the dynamical matrix
develops a plateau that extends to successively lower frequencies as
$\Delta \phi \rightarrow 0$~\cite{ohernJ} for both linear
and Hertzian springs (Eq.~\ref{spring}).  For intermediate and large
frequencies, the scaled $D({\overline \omega^d})$ does not depend
strongly on $\Delta \phi$.  However, we find that the weak peak in
$D({\overline \omega^d})$ diminishes and $D({\overline \omega^d})$
extends to slightly larger frequencies for Hertzian compared to linear
spring interactions.

\begin{figure}[h!]
\begin{center}
\includegraphics[width=0.7\textwidth]{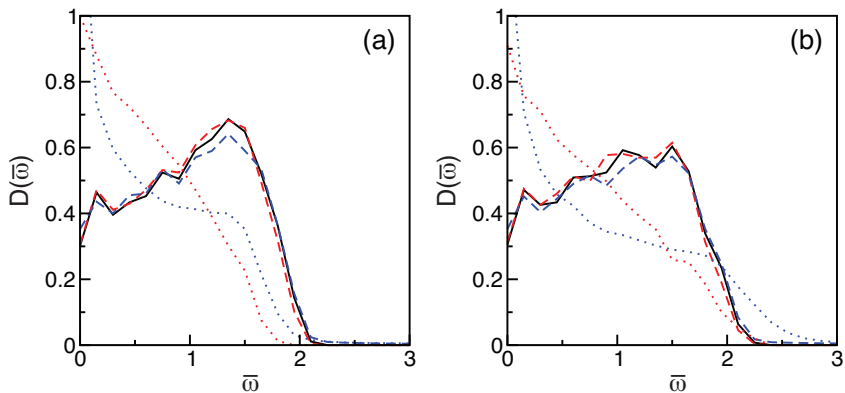}
\caption{The distribution of vibrational frequencies
$D({\overline\omega})$ for $N=32$ bidisperse disk packings with
$\Delta \phi=10^{-4}$ using the dynamical matrix (black), displacement
correlation matrix (red), and Fourier transform of the velocity
autocorrelation function (blue) at perturbation energies where $0\%$
(dashed) and $50\%$ (dotted) of the contacts are missing on average
from the zero-temperature configuration for purely repulsive (a) linear and
(b) Hertzian spring interactions.}
\label{dos_hertz}
\end{center}
\end{figure}

In Fig.~\ref{dos_hertz}, we show the binned distribution of
vibrational frequencies $D(\overline\omega)$ from the dynamical matrix
($D(\overline\omega^d)$), displacement correlation matrix
($D(\overline\omega^c)$), and Fourier transform of the velocity
autocorrelation function ($D(\overline\omega^v)$) for packings near
jamming onset with linear and Hertzian spring interactions as a function of the
perturbation energy.  At sufficiently low perturbation energies,
$D(\overline\omega^v)$ and $D(\overline\omega^c)$ both agree with
$D(\overline\omega^d)$. For both linear and Hertzian spring interactions,
$D(\overline\omega^v)$ and $D(\overline\omega^c)$ decay monotonically
from a peak near ${\overline \omega}=0$ at high perturbation energies
that remove on average $\approx 50\%$ of the original contacts that were
present at zero temperature.  However, $D(\overline\omega^v)$ and
$D(\overline \omega^c)$ do not have the same form at these energies;
$D(\overline\omega^v)$ possesses a plateau at intermediate frequencies,
whereas $D(\overline\omega^c)$ does not.  In addition, for large
perturbation energies, $D(\overline\omega^v)$ possesses a high
frequency tail for purely repulsive Hertzian compared to linear spring
interactions.

\begin{figure}[h!]
\begin{center}
\includegraphics[width=0.9\textwidth]{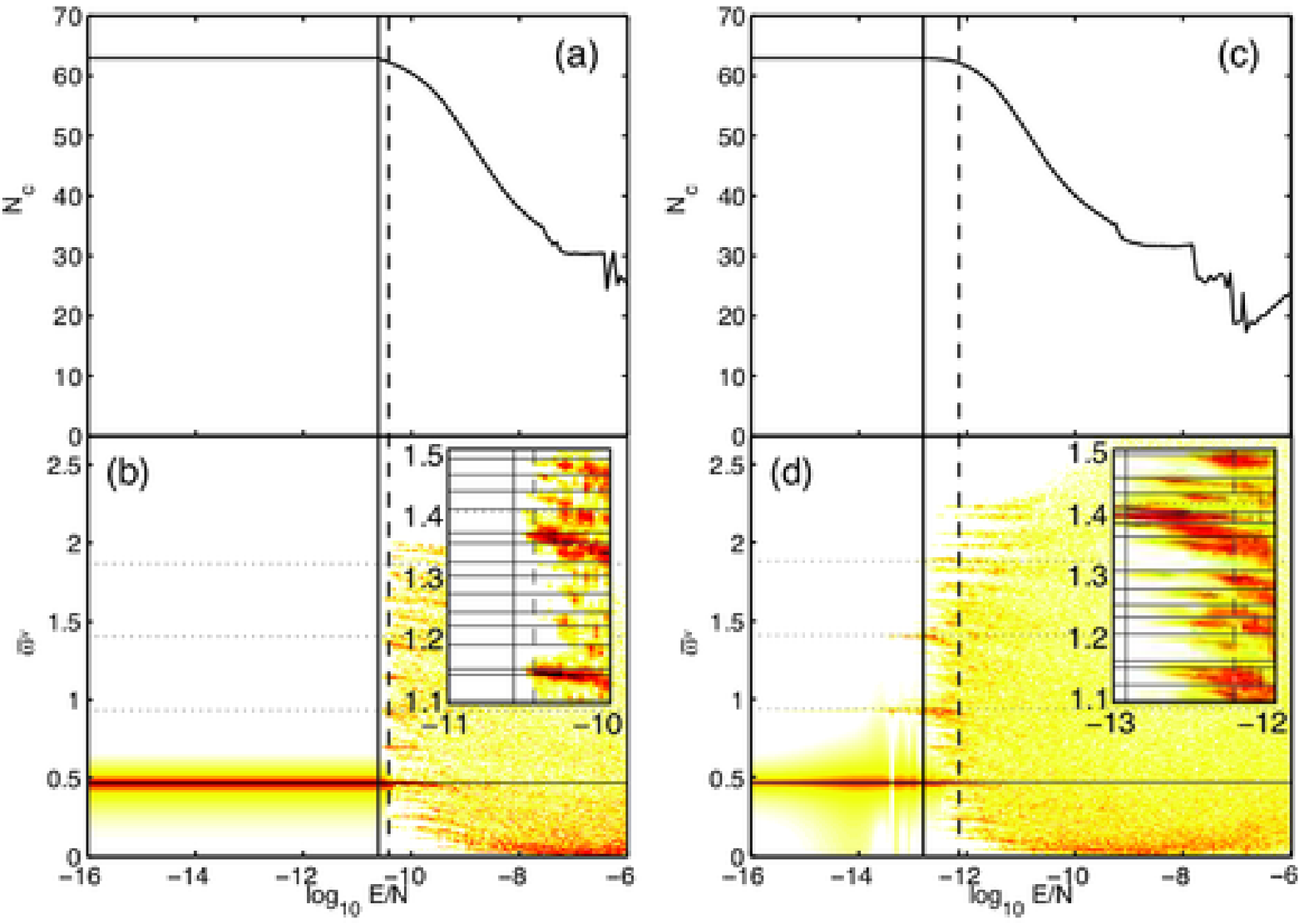}
\caption{Time-averaged number of contacts $N_c$ ((a) and (c)) and
color-scale plot of $\log_{10} D(\overline \omega^v)$ ((b) and (d))
for static packings of $N=32$ bidisperse disks perturbed along a
single eigenmode (mode $12$) as a function of the perturbation energy
$E/N$. At $E=0$, the packing possesses the isostatic number of
contacts $2N-1=63$ contacts. In (b) and (d), the solid horizontal line
represents the frequency of the driving frequency $\omega_{12}$ and
the dotted horizontal lines indicate harmonics of the driving
frequency, $2\omega_{12}$, $3\omega_{12}$, and $4\omega_{12}$.  The
vertical solid and dashed lines indicate the energies $E_c$ above
which the first contact breaks and $E_{1}$ above which there is on
average one contact missing from the zero-temperature
configuration. The inset shows a close-up of the region between
$E_c/N$ and $E_1/N$, where the solid horizontal lines give the
dynamical matrix frequencies. The left (right) columns show the
results for purely repulsive linear (Hertzian) spring interactions.}
\label{FT}
\end{center}
\end{figure}

In Fig.~\ref{FT}, we show the measured vibrational response (frequency
content of the Fourier transform of the velocity autocorrelation
function) for packings near jamming onset with $N=32$ that are
perturbed along eigenmode $12$ of the dynamical matrix over a range of
perturbation energies. At perturbation energies $E<E_c$, where $E_c$
is the energy above which a single contact breaks (Fig.~\ref{FT} (a)
and (c)), the vibrational response is confined to the original
eigenfrequency of the perturbation $\omega_{12}$ for repulsive linear
spring interactions (Fig.~\ref{FT} (b)).  In contrast, for packings
with purely repulsive Hertzian spring interactions~\cite{vakakis}, the
frequency response includes harmonics of the driving frequency
$\omega_{12}$ even before contact breaking (Fig.~\ref{FT} (d)), which
arise from the factor $(1-r_{ij}/\sigma_{ij})^{3/2}$ in the force law
(Eq.~\ref{spring}).

For perturbation energies beyond which a single contact breaks, but
before one contact is absent on average, $E_c<E<E_{1}$, the
vibrational response is described mainly by a set of discrete
frequencies. For both purely repulsive linear and Hertzian spring
interactions, the discrete frequencies correspond to a combination of
the eigenfrequencies of the dynamical matrix and low-frequency
harmonics of the driving frequency (although the low-frequency
harmonics possess a stronger signal for Hertzian spring interactions).  

In the Appendix, we demonstrate for a simpler two degree-of-freedom
system with double-sided Hertzian interactions that beats as well as
harmonics of the driving frequency are present at large perturbation
energies.  However, in packings with single-sided interactions,
contact breaking causes the vibrational response to become continuous
before a significant number of harmonics and beats occur. We find in
Fig.~\ref{FT} (b) and (d) that for $E>E_{1}$ the vibrational
response is described by nearly a uniform continuum of frequencies for
packings with both purely repulsive linear and Hertzian spring
interactions. Further, by comparing panels (a) and (b) (or (c) and (d)),
the vibrational response develops significant weight at zero frequency
when roughly $25\%$ of the zero-temperature contacts are missing on
average.  These results emphasize that contact breaking (not the shape
of the potential) dominates the nonlinear vibrational response for 
systems that interact via purely repulsive contact potentials~\cite{comment}. 

\begin{figure}[h!]
\begin{center}
\includegraphics[width=0.9\textwidth]{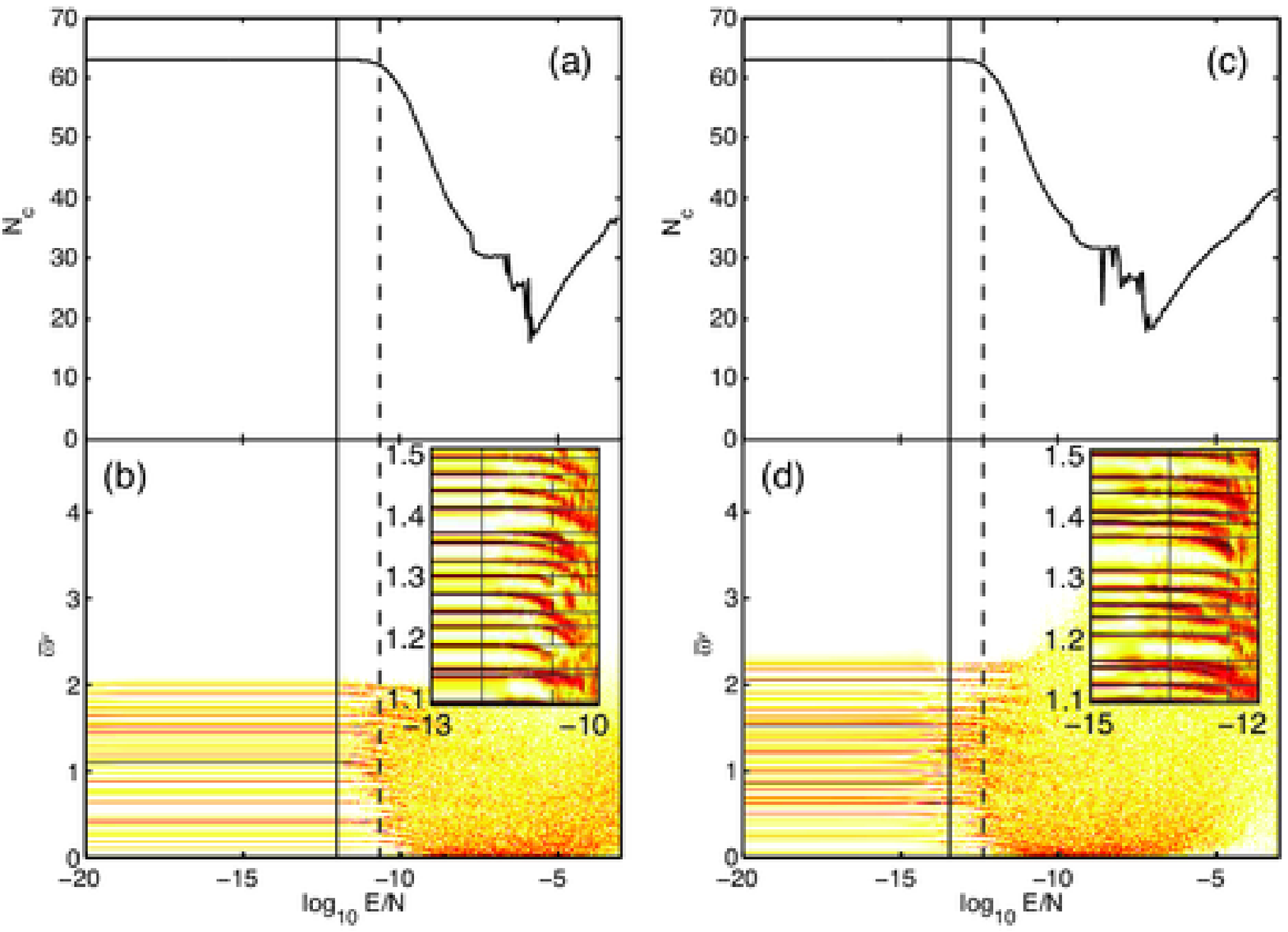}
\caption{Time-averaged number of contacts $N_c$ ((a) and (c)) and
color-scale plot of $\log_{10} D({\overline \omega^v})$ ((b) and (d))
for the same systems in Fig.~\ref{FT} perturbed in a superposition of
all eigenmodes of the dynamical matrix as a function of the
perturbation energy $E/N$. The vertical solid and dashed lines
indicate the energies $E_c$ above which the first contact breaks and
$E_{1}$ above which one contact on average is missing from the
zero-temperature configuration.  The insets are close-ups of the
vibrational response for $E_c < E < E_1$. The left (right) columns
show the results for purely repulsive linear (Hertzian) spring
interactions.}
\label{FT_all}
\end{center}
\end{figure}

How does the vibrational response depend on the nature of the
perturbation applied to the static packing?  In Fig.~\ref{FT_all} (b)
and (d), we show $D({\overline \omega^v})$ for the same packings
studied in Fig.~\ref{FT} after they are perturbed by a superposition
of eigenmodes $\vec{v}_{\rm all}$ instead of by a single eigenmode of
the dynamical matrix for both linear and Hertzian spring interactions.
Because we chose the perturbation $\vec{v}_{\rm all}$ to include equal
amounts of all dynamical matrix eigenmodes, at low perturbation
energies $E\ll E_c$, $D({\overline \omega^v})$ is composed of the
dynamical matrix eigenfrequencies with equal weights. At all
perturbation energies $E<E_c$, $D({\overline \omega^v})$ exhibits
peaks at the dynamical matrix eigenfrequencies with uniform weights
for linear spring interactions. In contrast, there are slight
differences in the weights at each of the dynamical matrix
eigenfrequencies for Hertzian interactions for $E \lesssim E_c$ and
for $E_c < E < E_1$ for both linear and Hertzian interactions.
Similar to the behavior when perturbing along a single eigenmode, we
find that the vibrational response includes a continuum of frequencies
for $E\gtrsim E_1$.  In addition, the density of vibrational modes
becomes peaked near zero frequency when the average number of contacts
drops to $\lesssim 25\%$ of its zero-temperature value.  In this regime, the
perturbation protocol does not affect the vibrational response.

\section{Conclusion}
\label{conclusion}

In summary, we performed molecular dynamics simulations to measure the
vibrational response of static disk packings near jamming onset that
interact via purely repulsive linear and Hertzian spring potentials.
Instead of assuming linear response as in many previous studies
(e.g. Refs.~\cite{ohernJ,liu}), we measured directly the vibrational
frequencies from the Fourier transform of the velocity autocorrelation
function and displacement correlation matrix.  We find that although
weak nonlinearities (e.g. the appearance of low-frequency harmonics of
the driving frequency) occur at energies below contact breaking for
packings that interact via Hertzian potentials, contact breaking
dominates the vibrational response for energies $E>E_1$. The onset of
nonlinearities from the shape of the Hertzian interaction potential
occurs at a similar energy to that for contact breaking, however,
contact breaking leads to much stronger nonlinearities in the
vibrational response, e.g. the response spreads to a continuum of
frequencies that are outside the range of the eigenfrequencies of the
dynamical matrix for the static packing. Thus, we have shown that
contact breaking gives rise to strongly nonharmonic response for
systems with both purely repulsive linear and Hertzian interaction
potentials.

\section{Acknowledgments}
\label{ack}

We acknowledge support from NSF Grant No. CBET-0968013 (MS), DTRA
Grant No. 1-10-1-0021 (CO), and Yale University (CS).
This work also benefited from the facilities and staff of the Yale
University Faculty of Arts and Sciences High Performance Computing
Center and the NSF (Grant No. CNS-0821132) that in part funded
acquisition of the computational facilities.  We thank Bob Behringer 
for his kind mentorship and honest science over the past 25 years.   

\appendix

\section{Simple Model}
\label{appendix1}

\begin{figure}[h!]
\begin{center}
\includegraphics[width=1\textwidth]{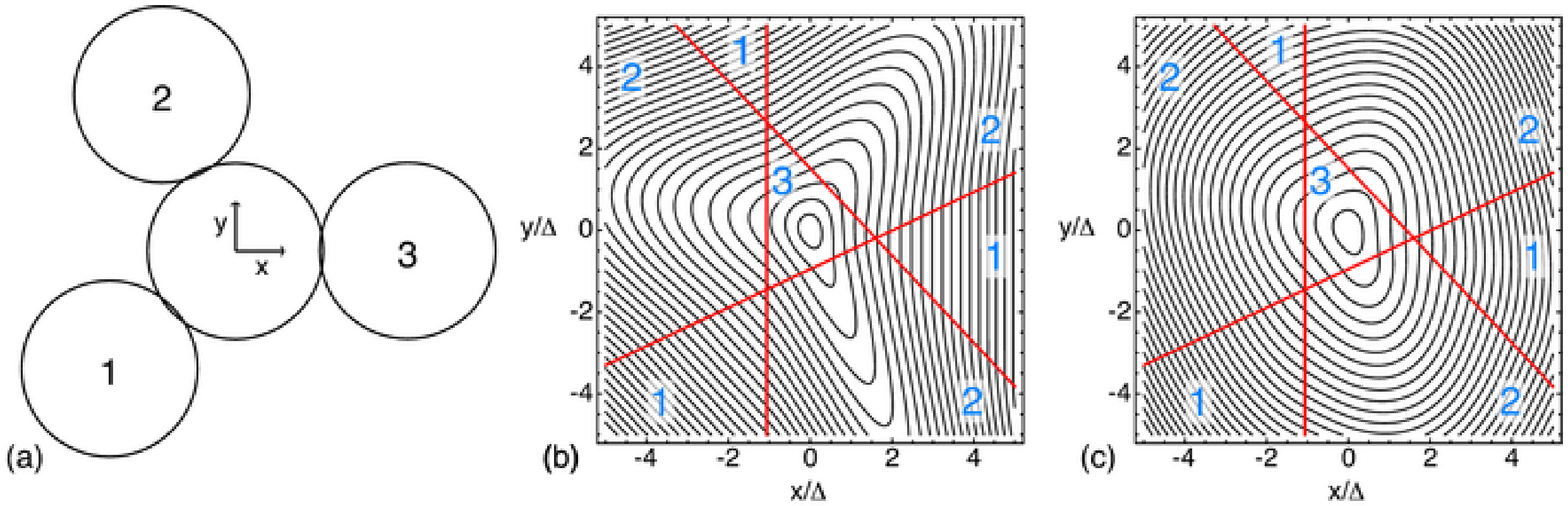}
\caption{Model system where a central mobile disk is confined by three
same-sized fixed disks $1$, $2$, and $3$ located at bond angles
$\theta_{12}=108^{\circ}$, $\theta_{23}=115.2^{\circ}$, and
$\theta_{31}=136.8^{\circ}$.  The initial amount of compression is
$\Delta=2\times10^{-2}$, and the disks interact via either single- or
double-sided Hertzian spring potentials.  (b) Contours of the
magnitude of the total force on the central disk $i$ from neighboring
disks $j$, $\big|\sum_{j=1}^3\vec{F}_{ij}\big|$, for single-sided
Hertzian interactions at compression $\Delta=10^{-4}$. The red lines
indicate when the central particle loses or gains a contact with its
neighbors, and the labels $1, 2,$ and $3$ correspond to the number of
contacts in each region bounded by the red lines. (c) Same as (b)
except the disks interact via double-sided Hertzian spring
interactions (Eq.~\ref{appspring}).}
\label{app_1}
\end{center}
\end{figure}

In this Appendix, we focus on a simple single-particle model to
illustrate that contact breaking rather than nonlinearities that arise
from the shape of the Hertzian potential dominates the vibrational
response of static packings with purely repulsive Hertzian spring
interactions.  This model consists of a central, mobile disk that is
confined between three fixed same-sized disks (Fig.~\ref{app_1}
(a)). The system is compressed from the `just-touching' configuration
by growing all particle diameters by $\sigma'=(1+\Delta)\sigma$, and
then using energy minimization to find the new mechanically stable
positions of the central disk $x'(\Delta)$ and $y'(\Delta)$.  The
disks interact via single-sided Hertzian spring forces
(Eq.~\ref{spring} with $\alpha=3/2$) or double-sided Hertzian spring
forces (Eq.~\ref{spring} with $\alpha=3/2$ without the Heaviside step
function)
\begin{eqnarray}
\label{appspring}
\vec{F}_{ij}&=&
\frac{\epsilon}{\sigma_{ij}}\bigg(1-\frac{r_{ij}}{\sigma}\bigg)^{3/2}\hat{r}_{ij},
\end{eqnarray}
for which all particles always interact with both repulsive ($r_{ij} <
\sigma$) and attractive ($r_{ij} > \sigma$) forces.  Studying this
simple system allows us to clearly separate the eigenfrequencies of the
dynamical matrix (linear response) from the additional frequencies in
the vibrational response that are generated from the shape of the
Hertzian potential and contact breaking (nonlinear response).

Figs.~\ref{app_1} (b) and (c) show the contours of the magnitude of
the total force on the central disk for single- and double-sided
Hertzian spring potentials, respectively.  When the force magnitude
contours are ellipsoidal and evenly spaced, the model system displays
linear response.  For both single- and double-sided Hertzian spring
potentials, we find that deviations from linear response occur at
energies near those that cause contact breaking (in single-sided
systems). When contacts break in systems with single-sided
interactions, the force magnitude contours become highly anisotropic
and multi-lobed. In contrast, the model system with double-sided
Hertzian spring interactions shows much weaker departures from linear
response at large energies where contacts would break in systems with 
single-sided interactions.

\begin{figure}[h!]
\begin{center}
\includegraphics[width=1\textwidth]{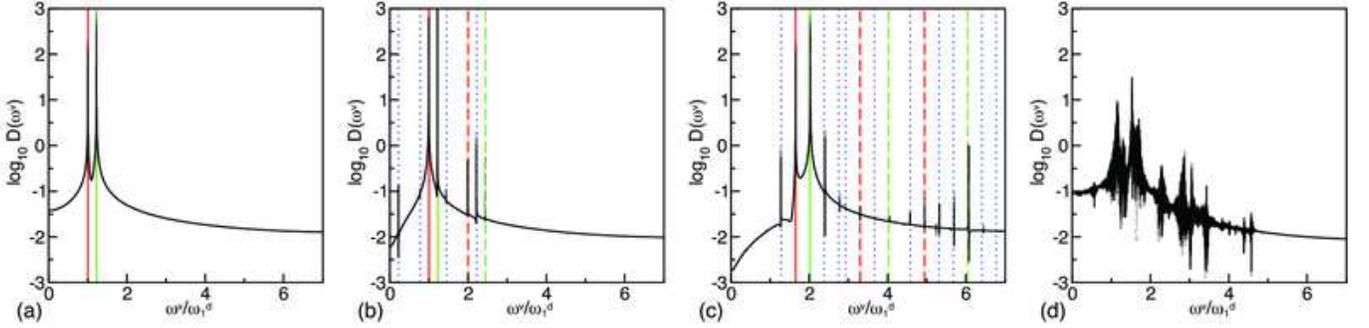}
\caption{Fourier transform of the velocity autocorrelation function
$D(\omega^v)$ for the model system in Fig.~\ref{app_1} (a) with
double-sided Hertzian spring interactions (Eq.~\ref{appspring})
compressed by $\Delta=10^{-4}$ and the central particle perturbed
randomly at energies (a) $E/N=10^{-16}$ (with an average number of
contacts $\langle N_c\rangle=3$), (b) $E/N=3\times10^{-11}$ just below
contact breaking for single-sided Hertzian spring interactions
($\langle N_c\rangle=3$), and (c) $E/N=10^{-7}$, which is far above
the energy required to break a single contact on average for
single-sided Hertzian spring potentials. In (d), we show the
vibrational response for perturbation energy $E/N=10^{-7}$ for the
same systems in (a)-(c) except the disks interact via single-sided
Hertzian spring potentials.  The vertical solid lines correspond to
the two dynamical matrix eigenfrequencies, the vertical dashed lines
correspond to harmonics of the dynamical matrix eigenfrequencies, and
the vertical dotted lines correspond to beats between the dynamical
matrix eigenfrequencies and their harmonics.  The frequencies
$\omega^v$ are normalized by $\omega^d_1$, which is the smallest
dynamical matrix eigenfrequency.  }
\label{app_2}
\end{center}
\end{figure}

In Fig.~\ref{app_2} (a)-(c), we show the vibrational response for the
model system with doubled-sided Hertzian spring interactions obtained
from the Fourier transform of the velocity autocorrelation function
for several values of the energy of the central particle.  At low
energies much below contact breaking (in systems with single-sided
interactions), only the two eigenfrequencies of the dynamical matrix
are found in the vibrational response (Fig.~\ref{app_2} (a)). For
energies close to and above contact breaking (in systems with
single-sided interactions), the eigenfrequencies of the dynamical
matrix, low-frequency harmonics, and beats between them are found in
the vibrational response (Fig.~\ref{app_2} (b) and (c)). In contrast,
at energies above contact breaking in packings with single-sided
Hertzian interactions the vibrational response shows a more continuous
spectrum of frequencies spread well beyond the range of the original
eigenfrequencies of the dynamical matrix. Previous work on Hertzian
chains with double-sided interactions has shown that the vibrational
response is periodic, but composed of non-sinusoidal nonlinear normal
modes~\cite{vak}.  Further, beats occur because these modes are not
orthogonal, and they mix with the eigenmodes of the dynamical matrix
and and their harmonics.  A detailed analysis of nonlinear normal 
modes in packings with single-sided Hertzian spring potentials will 
be pursued in future studies. 

\begin{thebibliography}{99}
\bibitem{foot}
Two of the eigenvectors of the dynamical matrix correspond to uniform
translations of the center of mass since we employ periodic boundary 
conditions in 2D.
\end{thebibliography}
\end{document}